\documentclass[letterpaper,11pt,conference]{IEEEtran}
\IEEEoverridecommandlockouts

\usepackage{cite}
\usepackage{amsmath,amssymb,amsfonts}
\usepackage{algorithmic}
\usepackage{graphicx}
\usepackage{textcomp}
\usepackage{hyperref}
\begin{document}

\title{\LARGE \bf
MEADE: Towards a Malicious Email Attachment Detection Engine
}

\author{
\IEEEauthorblockN{Ethan M. Rudd, Richard Harang, and Joshua Saxe}
\IEEEauthorblockA{
Sophos Group PLC\\
Data Science Division\\
Email: \{ethan.rudd,richard.harang,joshua.saxe\}@sophos.com\\
Address: Sophos PLC, 3975 University Drive Suite 330. Fairfax, VA 22030}}
\maketitle
\thispagestyle{empty}
\pagestyle{empty}

\begin{abstract}
Malicious email attachments are a growing delivery vector for malware. While machine learning has been successfully applied to portable executable (PE) malware detection, we ask, can we extend similar approaches to detect malware across heterogeneous file types commonly found in email attachments? In this paper, we explore the feasibility of applying machine learning as a static countermeasure to detect several types of malicious email attachments including Microsoft Office documents and Zip archives. To this end, we collected a dataset of over 5 million malicious/benign Microsoft Office documents from VirusTotal for evaluation as well as a dataset of benign Microsoft Office documents from the Common Crawl corpus, which we use to provide more realistic estimates of thresholds for false positive rates on in-the-wild data. We also collected a dataset of approximately 500k malicious/benign Zip archives, which we scraped using the VirusTotal service, on which we performed a separate evaluation. We analyze predictive performance of several classifiers on each of the VirusTotal datasets using a 70/30 train/test split on first seen time, evaluating feature and classifier types that have been applied successfully in commercial antimalware products and R\&D contexts. Using deep neural networks and gradient boosted decision trees, we are able to obtain ROC curves with $>$ 0.99 AUC on both Microsoft Office document and Zip archive datasets. Discussion of deployment viability in various antimalware contexts is provided.
\end{abstract}
\section{INTRODUCTION}

Email attachments are a straightforward delivery vector for malware campaigns, because, given a large enough organization, a sufficiently complacent user is bound to arise . 
Particularly in contexts when the contents of messages are impersonal and large files are exchanged on a routine basis (e.g., corporate environments), it is easy to provide attachments which look legitimate, yet contain malicious content\cite{wilding2017information}. 
Generally, attacks of this type require the user to download and/or open the attached file, and, from the adversary's perspective, it is often undesirable for the user to immediately recognize the file as malicious even after the payload gets executed -- ransomware, for example, takes time to index and encrypt targeted files\cite{kharraz2015cutting}. 
Thus, effective threat vectors, from an attacker's perspective are those that are commonly used by the targeted organization(s) yet have sufficient flexibility to both preserve legitimate looking content/structure and embed an attack.

In this paper, we examine two such types of attachments: Microsoft Office documents and Zip archives. Malicious Microsoft Office documents can be difficult to detect because they leverage ubiquitous functionalities that serve other purposes as well, for example, Microsoft Office documents allow embedding of multimedia, Visual Basic for Applications (VBA) macros, JavaScript, and even executable binaries to enhance functionality, usability, and aesthetics. These capabilities have led to high-quality office software that is easy-to-use, easy to augment, and aesthetically pleasing, by design, but that is also a vector for embedding malicious code. While this threat vector could easily be mitigated, e.g., by removing support for embedded VBA macros, such an approach is infeasible in practice since consumers of commercial software tend to favor functionality and aesthetics over security. Thus, When securing against Microsoft Office document vulnerabilities, we as security researchers and practitioners must walk the thin line between reducing consumer functionality on the one hand and mitigating the spread and execution of malware on the other. 

Archives (e.g., Zip, RAR) are even less constrained in the format of their internal contents than office documents, and can easily be packed internally with any file type. The inherent compression of most archive contents has led to their popularity for exchanging documents over email. However, an otherwise benign archive can easily be made malicious simply by insertion of one or more malicious files\cite{daryabar2011investigation}. In both malicious and benign settings, archives have been used to store code fragments that are later executed by software external to the archive, or conversely, archives have been embedded into other programs to form self-extracting archives\cite{ford2005vxa}. In a canonical malicious use-case, archives are distributed via phishing techniques \cite{noauthor_phishing_2014} such as impersonating an important contact, perhaps via spoofed email header, with the hope that the victim will unpack and run the archive's contents, e.g., a malicious JavaScript file executed outside of a browser sandbox. Such techniques have become increasingly common for malware propagation, particularly as of late.

Due to the unconstrained types of content that can be embedded into office documents and archive email attachments, they seem like natural candidates for machine learning. Unlike signature-based engines, machine learning offers the advantage that it can learn to generalize malicious behavior, and potentially generalize to new malware types. In this paper, we assess the viability of developing a machine-learnt static email attachment scanner for these file types, by leveraging techniques that have worked well for engines that detect other types of malware. Our work makes the following contributions:
\begin{itemize}
\item  We present evidence to support the viability of a static machine learning based email attachment detection.
\item We have collected datasets malicious/benign Microsoft Office documents and archives containing several million examples each.
\item We present evaluations on novel real-world attacks using Microsoft Office Documents infected with Petya, suggesting that ML methods may work when signature methods fail.
\item We present a more realistic evaluation method for conducting evaluations with noisy test data.
\item We present an evaluation of classifiers and feature types for office document and archive malware.
\end{itemize}
\section{Design Considerations and Related Work}

We were able to find little published work on malicious office document detection: Nissim et al. \cite{nissim2017aldocx} employ an approach which uses zip file paths across a small dataset of \textit{docx} files. However, their approach Open-XML specific and the dataset that they use is so small that it is difficult to draw conclusions about how their approach might perform in a realistic deployment scenario. Lagadec enumerates security issues with office document formats and how to manually detect them in \cite{lagadec2008opendocument}, but his approach does not utilize machine learning. While Lagadec \cite{lagadec2008opendocument} briefly discusses zip malware, we were only able to find peripheral discussions of archive malware.  Malicious PDF detection \cite{tzermias2011combining,laskov2011static} is a similar vein of research, but existing approaches focus on parsing the PDF format which is fragile and does not generalize well to other formats, and those that rely on dynamic analysis are expensive and do not scale nicely. We seek a generic approach that scales well.

While there are many academic papers that discuss the machine learning antimalware approaches, most use outdated and unrealistic datasets and do not map directly to real-world problems \cite{cruz2017open}. The classifier and feature types that we use in this paper were chosen based on techniques that have worked well for real-world antimalware problems involving detection of malicious PE files. Saxe and Berlin employ a deep neural network similar to ours \cite{saxe2015deep}, while Anderson and Roth employ gradient boosted decision trees \cite{anderson2018ember}. However, document and archive file formats have their own unique challenges because they are specifically designed to store user provided content which may or may not be executed, while PE files contain specified streams of execution.

As classifiers, we use feed-forward deep neural networks and gradient boosted decision ensembles. While one could try more sophisticated types of neural networks -- e.g., convolutional and recurrent, these are difficult to implement in practice due to large file sizes, computational overhead, and a dearth of generic byte-level embeddings. Though character-level embeddings have yielded success for certain antimalware problems, e.g., \cite{saxe2017expose}, these do not yet work well for generic byte-level embeddings of arbitrary length to our knowledge.  Thus, we instead transform each document/archive to a fixed-length feature vector before using it to train a classifier. Finally, we note that our focus in this paper is on on static detection, because machine learning models require a lot of data in order to work well. While antimalware stacks consist of both static and dynamic components, dynamic detection is very expensive computationally and is often employed to post-process detections from static engines, which operates much faster at scale. Dynamic detection is an important, complementary, and orthogonal area of research to that presented in this paper.

\section{File Structures}

Modern office documents generally fall into one of two types: the OLE2 standard \cite{noauthor_[ms-oleds]:_nodate} and the newer XML standard\cite{noauthor_introducing_nodate}. 
Microsoft Office's Word, Excel, and Powerpoint programs, along with analogous open source programs typically save OLE2 standard documents with .doc, .xls, and .ppt extensions and XML standard documents with .docx, .xlsx, and .pptx extensions. 
 The OLE2 standard was set forth by Microsoft and is also known as the Compound File Binary Format or Common Document File Format. 
OLE2 documents can be viewed as their own file-systems, analogous to FAT, wherein embedded streams are accessed via an index table. 
These streams can be viewed as sub-files and contain text, VBA macros, JavaScript, formatting objects, images, and even executable binary code.

Open XML formatted office documents contain similar objects, but are compressed as archives via Zip standard compression. 
Within each archive, the path to the embedded content is specified via XML. The user interface unpacks and renders relevant content within the Zip archive. Although the file format is different from OLE2, the types of embedded content contained are similar between the two formats. 

Open XML office documents are thus special cases of Zip archives, with a grounded well-defined structure, and in fact many file/archive types are special cases of the Zip format, including Java Archives (JARs), Android packages (APKs), and browser extensions.

\begin{figure}[h]
\centering
\includegraphics[height=3in]{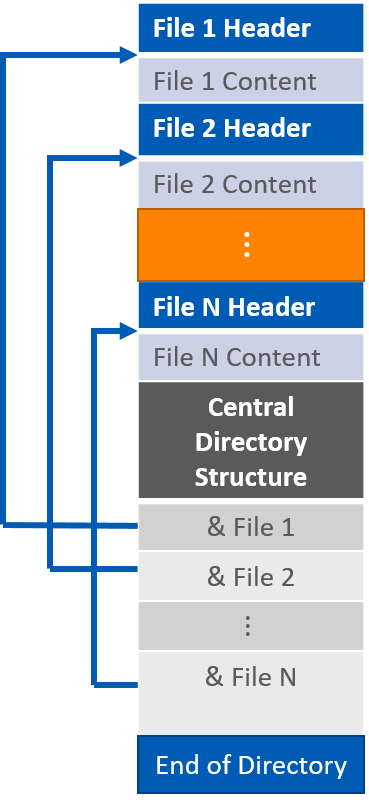} \hspace{0.2in} \includegraphics[height=3in]{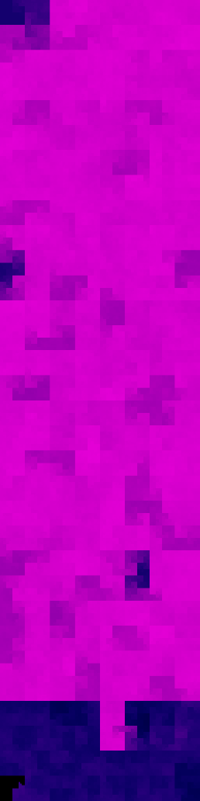}
\caption{Left: Structure of a Zip archive. The central directory structure resides just before the end of directory structure at the end of the archive. It contains an index of filenames and relative addresses within the archive. File headers contain additional metadata and are stacked above files which are compressed. Right: An entropy heat map of a zip archive plotted over a Hilbert Curve, generated using the BinVis \url{http://binvis.io/\#/} tool. The high-entropy regions (magenta) correspond to file contents, while the lower-entropy regions (blue/black) correspond to metadata. One can see that this archive contains three files, and one can easily discern the central directory structure at the end. Best viewed in color. }
\label{fig:zipfile}
\end{figure}

The structure of a Zip archive is shown in Fig. \ref{fig:zipfile}. The central directory structure, located near the end of the file contains names, references, and metadata about relevant files residing in the archive. The references in the central directory structure point to file headers, which contain additional metadata, followed by compressed versions of the files. 

\section{Feature Types}
\label{sec:features}

In order to train the classifiers specified in Sec. \ref{sec:classifiers}, fixed-size floating point vector representations of fields from input files/archives are required. From a practical perspective, these feature space representations must be reasonably efficient to extract,  particularly for archives, which can be hundreds of gigabytes in length. While we use concatenations of features extracted from different fields of files in our experiments, in this section, we describe the methods that we use to extract features from an arbitrary sequence of bytes. Several are adapted from \cite{saxe2015deep}. While the features that we use have been applied to other antimalware applications, we are the first to explore their utility for malicious email attachment detection.

{\bf $N$-gram Histograms} are derived from taking  $N$-gram frequencies \cite{rudd2017survey} over raw bytes. In practice, we use 3,4,5, and 6-grams, and apply the hashing trick \cite{chen2015compressing} to fix the dimensionality of the input feature space.

{\bf String Length-Hash Features} are obtained by applying delimiters to a sequence of bytes and taking frequency histograms of strings. Along a similar vein to Saxe and Berlin  \cite{saxe2017expose,saxe2015deep}, we apply the hashing trick over multiple logarithmic scales on string length and concatenate the resultant histograms into a fixed-size vector.

{\bf Byte Entropy Features} are obtained by taking a fixed-size sliding window, with a given stride, over a sequence of bytes and computing the entropy of each window. For each byte value, for a given window, the byte entropy calculation in that window (or zero) is stored, and a 2D histogram is taken over (byte value, entropy) pairs. The rasterized histogram becomes the fixed-size feature vector. In this paper, we employ a window size of 1024 with a stride of 256.

{\bf Byte Mean-Standard Deviation Features} are obtained using a similar fixed-size sliding window of given stride, but this time, the 2D histogram is taken over pairs of (byte mean, byte standard deviation) within each window.  The rasterized histogram becomes the fixed-size feature vector. Similar to byte entropy features, we we employ a window size of 1024 with a stride of 256.

\section{Classifier Types}
\label{sec:classifiers}

We conducted our evaluations using two classifier types that have achieved popularity for other antimalware tasks: deep neural networks (DNNs) and gradient boosted decision tree ensembles. While these classifiers are highly expressive and have advanced the state of the art in several problem domains, their formulations are quite different from one another. 

Neural networks consist of functional compositions of \textit{layers}, which map input vectors to output labels. The deeper the network, i.e., the more layers, the more expressive the composition, but also the greater the likelihood of over-fitting. Neural networks with more than one hidden (non input or output) layer are said to be ``deep neural networks''.  In our case, the input vector is a numerical representation of bytes from a file, and the output is a scalar malicious or benign label. The (vector,label) pairs are provided during training for the model to learn the parameters of the composition. Further details on DNNs can be found in \cite{goodfellow2016deep}.  We implemented our DNN in \textit{Keras}\cite{chollet2015keras}, employing a similar topology to that of Saxe and Berlin\cite{saxe2015deep}, using 4 hidden layers of size 1024 each with rectified linear unit (ReLU) activations. At each layer we employ dropout and and batch normalization regularization methods, with a dropout ratio of 0.2. At the final output we use a sigmoid cross-entropy loss function: 

\begin{equation}
\footnotesize
\label{eq:sigmoid_cost}
J(x_i;y_i,\theta) =  y_i log\sigma(f(x_i);\theta) + (1-y_i)log(1-\sigma(f(x_i);\theta),
\end{equation}

\noindent
where $\theta$ correspond to all parameters over the network, $x_i$ corresponds to the $i$th training example, $y_i$ corresponds to the label for that example, $f(x_i)$ corresponds to the pre-activation output of the final layer, and $\sigma(\cdot)$ is the logistic sigmoid function. We optimized $\theta$ using the Keras framework's default \textit{ADAM} \cite{kingma2014adam} solver, with minibatch size of 10k, and performed early stopping when loss over a validation set failed to decrease for 10 consecutive epochs. 

Decision trees, instead of trying to learn a latent representation whereby data separates linearly, partition the input feature space directly in a piecewise-linear manner. While they can fit extremely nonlinear datasets, the resultant decision boundaries also tend to exhibit extremely high variance. By aggregating an ensemble of trees, this variance can be decreased. Gradient boosting \cite{ye2009stochastic} iteratively adds trees to the ensemble; given loss function $J(F(x;\theta),y)$, and classification function $F(x;\theta)$  for the ensemble, a subsequent tree is added to the ensemble at each iteration to fit pseudo-residuals of the training set, $-\frac{\partial J(F(x_i;\theta),y_i)}{\partial F(x_i;\theta)}$. The subsequent tree's decisions are then weighted so as to minimize the loss of the overall ensemble. For our gradient boosted ensembles, we used a regularized logistic sigmoid cross-entropy loss function, similar to that of our neural network (cf. Eq. \ref{eq:sigmoid_cost}), but unlike with the network, wherein all parameters are jointly optimized with respect to the cost function, the ensemble is iteratively refined with the addition of each decision tree -- i.e., additional parameters are added to the model. Our implementation uses the eXtreme Gradient Boosting (\textit{XGBoost}) implementation by Chen et al. \cite{chen2016xgboost}. For our choice of hyperparameters, we employed a maximum depth per tree of 6, a subsample ratio of 0.5 (on training data; not columns), and hyperparameter $\eta$ of 0.1. We used ten rounds without improvement in classification accuracy over a validation set as a stopping criterion for growing the ensemble.

\section{Office Documents: Experiments and Discussion}
\label{sec:Documents}

\begin{figure}[!b]
\centering
\includegraphics[width=0.5\textwidth]{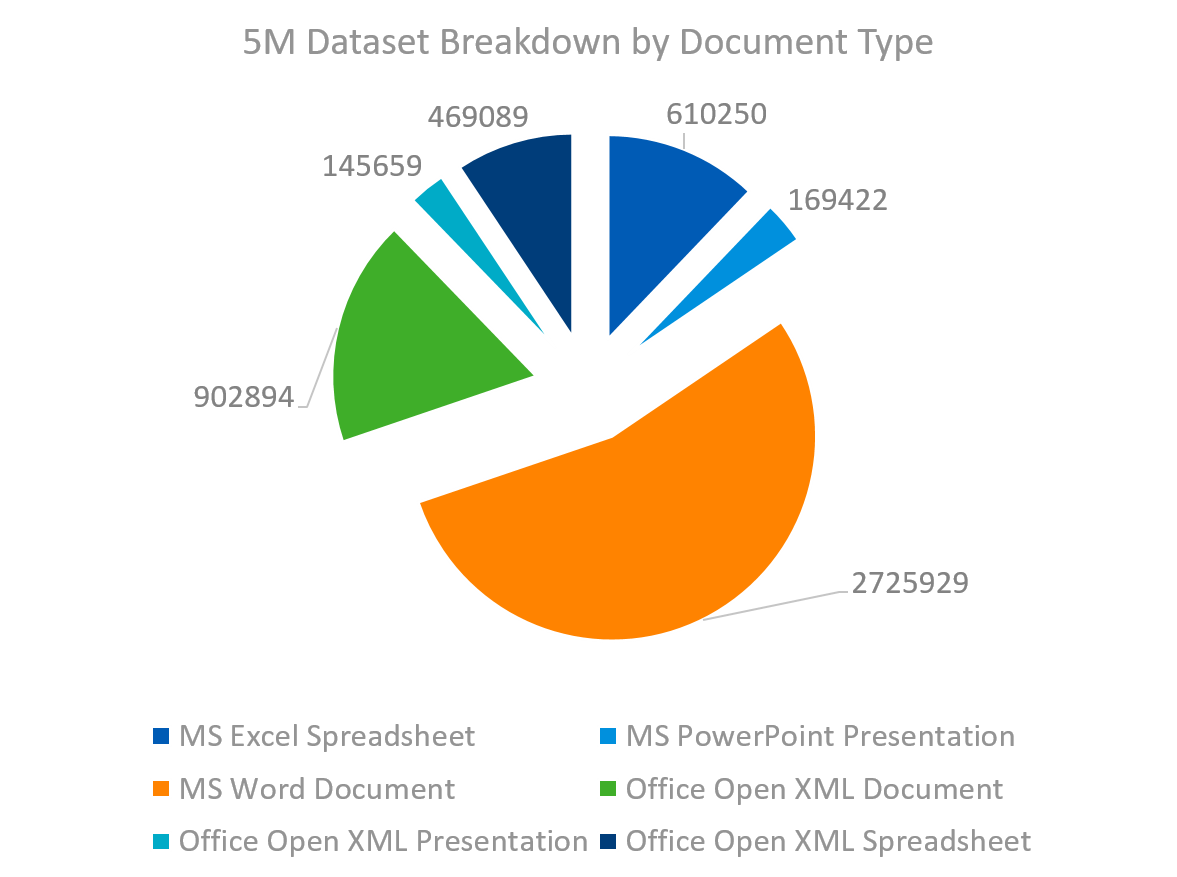}
\caption{Numeric breakdown of VirusTotal dataset documents by file type. The majority of available data consists of legacy (.doc) and new (.docx) word processing formats. Best viewed in color.
}
\label{fig:5MDataDist}
\end{figure}

We collected an initial dataset of 5,023,243 malicious and benign office documents by scraping files and reports from the VirusTotal \cite{virustotal} service, a service which submits files to a variety of antimalware products and returns vendor responses. We assign malicious/benign labels on a 5+/1- basis, i.e., documents for which one or fewer vendors labeled malicious, we ascribe the aggregate label \textit{benign}, while documents for which 5 or more vendors labeled malicious, we ascribe the aggregate label \textit{malicious}. We initially collected 6 million documents, but leave those with between 2 and 4 (inclusive) vendor responses out of our dataset. We use this 5+/1-  criterion in part because vendor label information is given after some time lag between the first seen time of a sample and the last scan time of the sample, and when we deploy we want our classifier to be able to make a good prediction that is somewhat unaffected by biases within the vendor community. Empirical analysis internal to Sophos suggests that this labeling scheme works reasonably well for assigning aggregate malicious/benign scores. Note also that due to the fundamental generalized nature of machine learning, the goal here is not merely to emulate vendor aggregation but to learn predictive latent patterns that correctly make future predictions of malicious/benign when other vendors' signature-based methods fail. The breakdown of VirusTotal derived document dataset by format type is shown in Fig. \ref{fig:5MDataDist}.

Since our objective is to obtain a predictive classifier, we performed a 70/30 quantile split on the \textit{first\_seen} timestamp from VirusTotal, allocating the first 70th percentile as a training set and the last 30th percentile as our test set. Note that for realistic evaluation, training samples must come temporally before test samples. As malware is constantly evolving, simply performing a $k$-fold cross validation across all data is unrealistic because it implicitly assumes that we have information about future malware samples, and would thus lead to a grossly inflated estimate of the efficacy of the classifier.  To obtain an estimate of the \emph{deployment performance} of the classifier, where it will be trained on available data and then used to detect (potentially novel) malware from that point forward, we must ensure that the evaluation set is composed of files obtained after the newest file in the training set.

We conducted a multitude of experiments using both DNN and XGBoost classifiers with byte entropy histograms, string length-hash histograms, and byte mean-standard deviation histograms as features.  We extracted features across whole documents, and found that length-hash features disproportionately performed the best of any one feature type when delimiting by non-printable characters as well as  ``$<$'', ``$>$'', ``$/$'', ``\textbackslash'', and `` ''. Byte entropy and mean-standard deviation histograms were uniformly spaced along each axis, initially to have a total of 1024 bins, then later downsized to 256 bins each after experiments indicated negligible gain from added feature dimension. String length-hash features were configured to have a total of 1024 dimensions; 64 per logarithmic scale of string length. Only strings between 5 and 128 characters were considered, with the remainder ignored. We also logarithmically scaled the bins as we found that this resulted in a slight performance increase. We also tried unpacking and concatenating the contents of compressed Open XML format documents prior to extracting features. Surprisingly, this resulted in a performance decrease, which suggests that our classifiers predominantly learn from file metadata.

\begin{figure}[!t]
\centering
\includegraphics[width=0.5\textwidth]{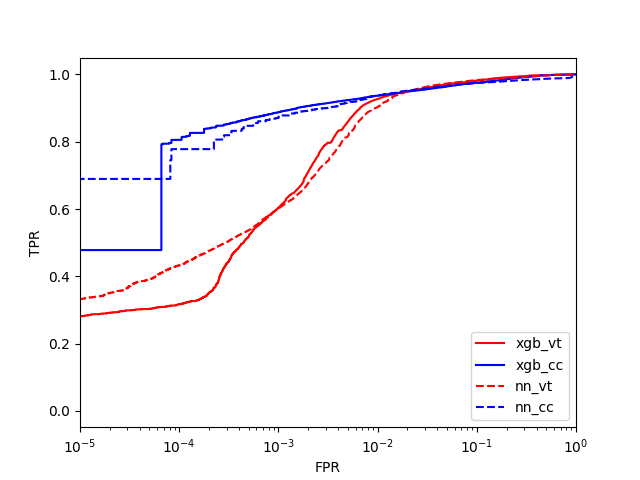}
\caption{ROC curves for office document DNN and XGB classifiers with FPR/thresholds assessed over the VirusTotal dataset (VT) and Common Crawl dataset (CC). We believe that the VirusTotal dataset is rife with false positives that all vendors miss since files submitted are disproportionately malicious/suspicious, and that obtaining TPRs on the VT dataset at FPRs/thresholds derived from the CC dataset yields a more realistic evaluation.}
\label{fig:docroc}
\end{figure}

Using concatenations of all feature vectors -- string length-hash, byte entropy, and byte mean-standard deviation histograms, for both DNN and XGBoost Classifiers, we were able to obtain an area under a receiver operating characteristics (ROC) curve of greater than 0.99, with the DNN slightly outperforming XGBoost (cf. red lines in Fig. \ref{fig:docroc}). Using the same features to train a linear support vector machine under a tuned C value yielded less than 0.90 AUC, suggesting that expressive nonlinear concepts can indeed be derived from our input feature space representations, pointing to the utility of more expressive nonlinear classifiers. 

Interestingly, we found that the DNN's performance did not noticeably improve when using a concatenation of all features, as opposed to just string length-hash features, but XGBoost's performance improved substantially. This suggests that our DNN architecture is favorable, from a deployment perspective, as feature extraction accounts for the majority of processing time at inference -- particularly when classifying large documents. 

As an exploratory analysis, we also tried using the outputs from intermediate layers of the trained network on our train and test sets as feature vectors for XGBoost, since the learning processes of the two classifiers are fundamentally different, but this resulted in a performance degradation. We also tried training models with additional hidden layers, which yielded slightly decreased performance, as well as separate malicious/benign outputs -- one per file-type -- along with a global malicious/benign score under a MOON-like topology \cite{rudd2016moon}. While the MOON-like network yielded slightly better performance in low FPR regions of the ROC, performance deteriorated in higher FPR regions, yielding no net gains for the added complexity.

During our evaluation, we conducted our own forensic investigation of the dataset, dumping VBA macros for 100 ``benign" files from VirusTotal that our DNN labeled malicious with high confidence. In the majority of cases, we found signs of malicious payloads and code obfuscation, suggesting that a good number of ``false positives'' from our VirusTotal dataset might actually be false negative novel attacks that all vendors missed.

\begin{figure}[!t]
\centering
\includegraphics[width=0.5\textwidth]{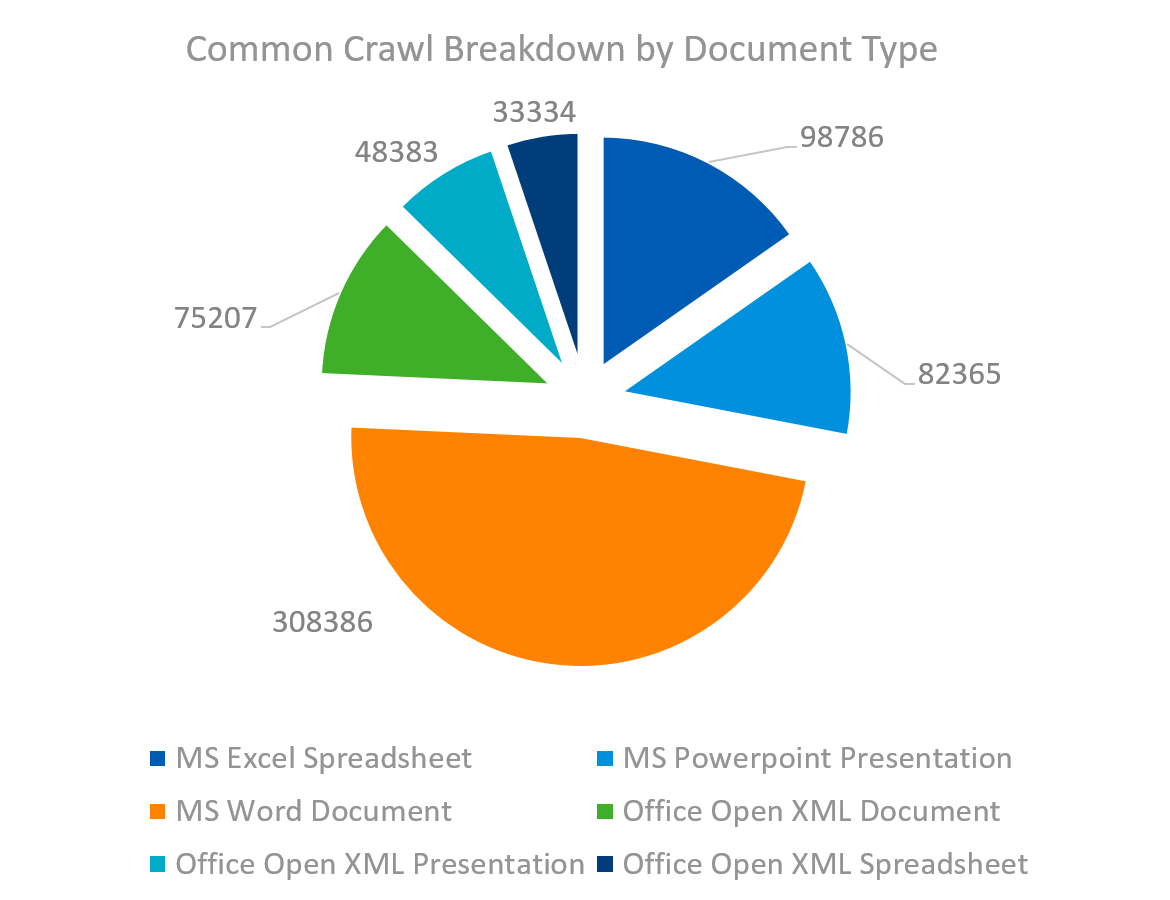}
\caption{Numeric breakdown of Common Crawl dataset documents by file type. As with the VirusTotal datset, the majority of available data consists of legacy (.doc) and new (.docx) word processing formats, suggesting coarse alignment in terms of dataset balance/bias. This is an important consideration when using the dataset to assess realistic thresholds for false positive rate. Best viewed in color.}
\label{fig:CCFDist}
\end{figure}

This finding from our forensic analysis leads us to consider that using vendor labels from VirusTotal as a test criterion may be implicitly biasing our success criteria to currently existing classifiers and unfairly penalizing those capable of generalizing to novel malicious behavior seen in novel attacks. After all, most of the vendor scores in VT come from signature-based -- not machine learnt -- anti-malware engines. Thus, we surmise that using VT gives us an unfairly pessimistic estimate of false positive rate. This is exacerbated because files submitted to VT are far more likely to be malicious (or at least suspicious) than most files in the wild.

We therefore collected an additional corpus of approximately 1 million likely benign documents scraped from known benign URLs from Common Crawl \cite{commoncrawl} -- a web archiving service -- and submitted these to VirusTotal for labeling. Of the Common Crawl documents, only a 15 were labeled as malicious. Discarding these and taking the rest as benign, we used this Common Crawl-derived dataset to re-evaluate false positive rate, and using corresponding thresholds, estimated the true positive rate on VirusTotal. Via this procedure, we were able to achieve noticeable gains (cf. the blue lines in Fig.\ref{fig:docroc}). Note that this may even be an under-estimate of true performance because gains in the network from detecting mislabeled false negatives in our VirusTotal dataset are not recognized (but at least now they are not penalized).

As an additional qualitative analysis of our network's capability to generalize malicious concepts, we conducted an analysis on office documents infected by the recent Petya ransomware, a malware notorious for employing novel exploits\cite{goodin_notpetya_2017}. While we will not comment on specific vendors' detection capabilities, Petya was able to propagate undetected and cause a global cyber security crisis despite the presence of numerous antimalware engines. At a threshold yielding an FPR of 1e-3 assessed on our Common Crawl dataset, we were able to detect 5 out of 9 malicious Petya samples, which provides further (albeit anecdotal) evidence that our DNN may have learned generalized malicious concepts within its latent representations beyond any capacity of signature-driven systems. Note also that all data upon which the network was trained was collected prior to the Petya outbreak. 

\section{Zip Archives: Experiments and Discussion}
\label{sec:Archive}

Along a similar vein to our office document dataset, we collected a dataset of approximately 500k Zip archives by scraping VirusTotal. We found that Zip archives exhibited much larger variation in size than office documents. We performed a similar 70/30 train/test split on timestamps as we did for office documents, grouping samples with VirusTotal \textit{first\_seen} timestamps in the first 70th percentile into our training set and samples with \textit{first\_seen} timestamps in the last 30th percentile into our test set. 

While for such a small  dataset we could simply extract and concatenate content and metadata, from a practical perspective, this becomes problematic when dealing with large, potentially nested zip archives. Moreover, findings from Sec. \ref{sec:Documents} suggest that useful features for classification are typically contained within metadata for a very structured subset of Zip archives (Open XML format office documents). Extracting similar string length-hash features over entire archives and fitting a DNN, yielded an ROC with an AUC of less than 0.9, which is not useful for commercial antimalware applications. 

We hypothesized that this poor performance was due to a low signal-to-noise ratio in the feature space, and thus chose to extract a set of features over more relevant sections of Zip archives. Using our knowledge of Zip archive structure (cf. Fig. \ref{fig:zipfile}), we arrived at an easy-to-extract set of features: First, by matching  appropriate magic numbers, we dumped raw bytes from each archive's central directory structure. We then dumped the last 1 MB of the archive's raw bytes, or the entire archive for archives less than 1 MB in size. Over the central directory structures, we extracted 1024 dimensional feature vectors: string length-hash histograms, byte entropy features, and hashed 3,4,5, and 6 grams. Over the last 1 MB we extracted 1024 MB byte entropy features and string length-hash histograms. We omitted n-grams due to lengthy extraction times. For our string length-hash features, we used a similar parameterization to Sec. \ref{sec:Documents}, except that we used length 2 as a lower-bound cutoff for considering a given string. 

\begin{figure}[!t]
\centering
\includegraphics[width=0.5\textwidth]{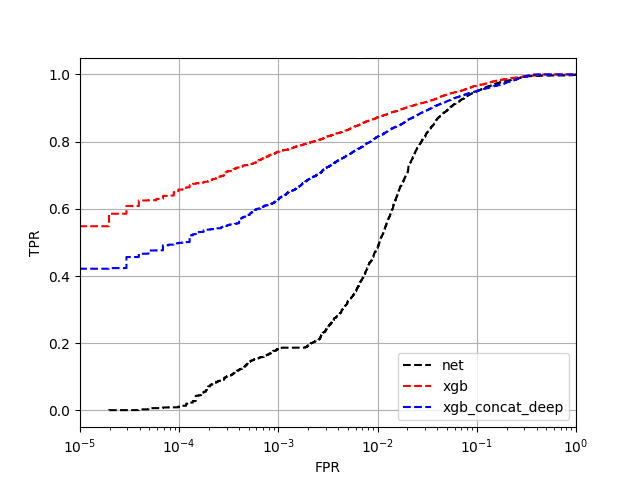}
\caption{ROC curves for the best DNN (red) and XGBoost (black) classifiers using the Zip archive dataset. The dashed blue line (middle) was obtained by concatenating deep features obtained from the network to the original feature vectors and performing training/testing over these using XGBoost classifier over these. The performance decay suggests that the DNN has not learned a good feature space representation for the problem at hand.}
\label{fig:ArchiveROC}
\end{figure}

As classifiers, we used the same XGBoost and DNN classifiers described in Secs. \ref{sec:Archive} and \ref{sec:Documents} Results are shown in Fig. \ref{fig:ArchiveROC}. Surprisingly, the DNN's performance was inferior to XGB over any single feature type, and using a concatenated 5120-dimensional feature vector the network noticeably under-performed XGB, offering an ROC with an AUC of 0.98. Concatenating all features using XGBoost yielded an AUC of greater than 0.99, with differences particularly pronounced in low-FPR regions.

Via the same methodology in Sec. \ref{sec:Documents}, we used the network to extract deep features, concatenated them with the five feature types, and fit an XGBoost classifier. This resulted in noticeably diminished performance for the XGBoost classifier, suggesting that the network was unable to learn a meaningful feature space. Perhaps this problem could be ameliorated by using a larger archive dataset, or perhaps archives are intrinsically harder to learn meaningful features on than documents.

\section{Conclusions and Future Work}

The exploratory research conducted herein suggests that machine learning is a viable approach for certain malicious email attachment scanner applications, particularly those tuned for a high false positive rate, where false positives are passed to a secondary scanner for enhanced detection -- e.g., a dynamic detection engine in a sandbox. Using fixed-size histogram features as input, both DNN and XGB classifiers offered comparable performance for office document data, but XGB far and away outperformed DNNs on generic Zip archive data. Perhaps, a larger amount of data is simply required for DNNs to perform well. The fact that concatenating deep features as input to our XGB classifiers did not improve XGB suggests perhaps that our DNNs are simply learning to ``memorize'' interesting patterns without deriving feature spaces that offer smooth statistical support. In the future, we plan to collect a larger datasets and perhaps explore additional attachment types (e.g., RAR, 7ZIP, GZIP, CAB, PDF, etc.). Even if the network tends to use its capacity for ``memorization'', more data could render a deep learning approach practical with respect to accuracy. However, ``memorization" may pose a problem with respect to generalization capability and suggests potential  susceptibility to novel malware types or adversarial attacks. While an analysis of adversarial attacks is beyond the scope of this work, we note that our feature representation may make it difficult to evade our model with gradient-based attacks, such as those developed in \cite{rozsa2016adversarial,rozsa2016facial,rozsa2017facial,papernot2016limitations,carlini2017adversarial}. Such attacks produce modified feature vectors that are `close' to the original feature vector under some distance measure that are misclassified.  Because our feature representations make heavy use of hashing, identifying the modifications to a file that would produce the desired modifications in the feature vector would require, for at least some features, inverting the chosen hash function.

There is a lot of room to expand the use of deep learning both for malicious email attachment detection and other anti-malware applications that we and others in the community are researching, including learning embeddings and sequence models over features from different sections of a file, leveraging large quantities of unlabeled data, e.g., via ladder networks\cite{rasmus2015semi}, and discovery of generic byte-level embeddings. It will be interesting to see the viability of these approaches in the coming years.

From a deployment perspective, how a combined model might look across different attachment types is an interesting question. Given a centralized mail server, one option would be to have many classifiers for different attachment formats, but a better solution from an endpoint perspective would be to have a single shared representation handle multiple file types. The extent to which a joint representation can be leveraged over multiple file types and formats without introducing the problem of catastrophic forgetting\cite{french1999catastrophic} is an important area for future research.

\section*{ACKNOWLEDGMENT}

This work was funded by Sophos PLC.

\bibliographystyle{unsrt}
\bibliography{references}

\begin{thebibliography}{10}

\bibitem{wilding2017information}
Edward Wilding.
\newblock {\em Information risk and security: preventing and investigating
  workplace computer crime}.
\newblock Routledge, 2017.

\bibitem{kharraz2015cutting}
Amin Kharraz, William Robertson, Davide Balzarotti, Leyla Bilge, and Engin
  Kirda.
\newblock Cutting the gordian knot: A look under the hood of ransomware
  attacks.
\newblock In {\em International Conference on Detection of Intrusions and
  Malware, and Vulnerability Assessment}, pages 3--24. Springer, 2015.

\bibitem{daryabar2011investigation}
Farid Daryabar, Ali Dehghantanha, and Hoorang~Ghasem Broujerdi.
\newblock Investigation of malware defence and detection techniques.
\newblock {\em International Journal of Digital Information and Wireless
  Communications (IJDIWC)}, 1(3):645--650, 2011.

\bibitem{ford2005vxa}
Bryan Ford.
\newblock Vxa: A virtual architecture for durable compressed archives.
\newblock In {\em FAST}, volume~5, pages 295--308, 2005.

\bibitem{noauthor_phishing_2014}
Phishing with a malicious .zip attachment, April 2014.

\bibitem{nissim2017aldocx}
Nir Nissim, Aviad Cohen, and Yuval Elovici.
\newblock Aldocx: detection of unknown malicious microsoft office documents
  using designated active learning methods based on new structural feature
  extraction methodology.
\newblock {\em IEEE Transactions on Information Forensics and Security},
  12(3):631--646, 2017.

\bibitem{lagadec2008opendocument}
Philippe Lagadec.
\newblock Opendocument and open xml security (openoffice. org and ms office
  2007).
\newblock {\em Journal in Computer Virology}, 4(2):115--125, 2008.

\bibitem{tzermias2011combining}
Zacharias Tzermias, Giorgos Sykiotakis, Michalis Polychronakis, and Evangelos~P
  Markatos.
\newblock Combining static and dynamic analysis for the detection of malicious
  documents.
\newblock In {\em Proceedings of the Fourth European Workshop on System
  Security}, page~4. ACM, 2011.

\bibitem{laskov2011static}
Pavel Laskov and Nedim {\v{S}}rndi{\'c}.
\newblock Static detection of malicious javascript-bearing pdf documents.
\newblock In {\em Proceedings of the 27th annual computer security applications
  conference}, pages 373--382. ACM, 2011.

\bibitem{cruz2017open}
Steve Cruz, Cora Coleman, Ethan~M Rudd, and Terrance~E Boult.
\newblock Open set intrusion recognition for fine-grained attack
  categorization.
\newblock In {\em Technologies for Homeland Security (HST), 2017 IEEE
  International Symposium on}, pages 1--6. IEEE, 2017.

\bibitem{saxe2015deep}
Joshua Saxe and Konstantin Berlin.
\newblock Deep neural network based malware detection using two dimensional
  binary program features.
\newblock In {\em Malicious and Unwanted Software (MALWARE), 2015 10th
  International Conference on}, pages 11--20. IEEE, 2015.

\bibitem{anderson2018ember}
Hyrum~S Anderson and Phil Roth.
\newblock Ember: An open dataset for training static pe malware machine
  learning models.
\newblock {\em arXiv preprint arXiv:1804.04637}, 2018.

\bibitem{saxe2017expose}
Joshua Saxe and Konstantin Berlin.
\newblock expose: A character-level convolutional neural network with
  embeddings for detecting malicious urls, file paths and registry keys.
\newblock {\em arXiv preprint arXiv:1702.08568}, 2017.

\bibitem{noauthor_[ms-oleds]:_nodate}
[{MS}-{OLEDS}]: {OLE}1.0 and {OLE}2.0 {Formats}.

\bibitem{noauthor_introducing_nodate}
Introducing the {Office} (2007) {Open} {XML} {File} {Formats}.

\bibitem{rudd2017survey}
Ethan Rudd, Andras Rozsa, Manuel Gunther, and Terrance Boult.
\newblock A survey of stealth malware: Attacks, mitigation measures, and steps
  toward autonomous open world solutions.
\newblock {\em IEEE Communications Surveys \& Tutorials}, 2017.

\bibitem{chen2015compressing}
Wenlin Chen, James Wilson, Stephen Tyree, Kilian Weinberger, and Yixin Chen.
\newblock Compressing neural networks with the hashing trick.
\newblock In {\em International Conference on Machine Learning}, pages
  2285--2294, 2015.

\bibitem{goodfellow2016deep}
Ian Goodfellow, Yoshua Bengio, and Aaron Courville.
\newblock {\em Deep learning}.
\newblock MIT press, 2016.

\bibitem{chollet2015keras}
Fran{\c{c}}ois Chollet et~al.
\newblock Keras, 2015.

\bibitem{kingma2014adam}
Diederik Kingma and Jimmy Ba.
\newblock Adam: A method for stochastic optimization.
\newblock {\em arXiv preprint arXiv:1412.6980}, 2014.

\bibitem{ye2009stochastic}
Jerry Ye, Jyh-Herng Chow, Jiang Chen, and Zhaohui Zheng.
\newblock Stochastic gradient boosted distributed decision trees.
\newblock In {\em Proceedings of the 18th ACM conference on Information and
  knowledge management}, pages 2061--2064. ACM, 2009.

\bibitem{chen2016xgboost}
Tianqi Chen and Carlos Guestrin.
\newblock Xgboost: A scalable tree boosting system.
\newblock In {\em Proceedings of the 22nd acm sigkdd international conference
  on knowledge discovery and data mining}, pages 785--794. ACM, 2016.

\bibitem{virustotal}
Virus Total.
\newblock Virustotal-free online virus, malware and url scanner.
\newblock {\em Online: https://www. virustotal. com/en}, 2012.

\bibitem{rudd2016moon}
Ethan~M Rudd, Manuel G{\"u}nther, and Terrance~E Boult.
\newblock Moon: A mixed objective optimization network for the recognition of
  facial attributes.
\newblock In {\em European Conference on Computer Vision}, pages 19--35.
  Springer, 2016.

\bibitem{commoncrawl}
Jason~R Smith, Herve Saint-Amand, Magdalena Plamada, Philipp Koehn, Chris
  Callison-Burch, and Adam Lopez.
\newblock Dirt cheap web-scale parallel text from the common crawl.
\newblock In {\em ACL (1)}, pages 1374--1383, 2013.

\bibitem{goodin_notpetya_2017}
Dan Goodin.
\newblock {NotPetya} developers may have obtained {NSA} exploits weeks before
  their public leak [{Updated}], June 2017.

\bibitem{rozsa2016adversarial}
Andras Rozsa, Ethan~M Rudd, and Terrance~E Boult.
\newblock Adversarial diversity and hard positive generation.
\newblock In {\em Proceedings of the IEEE Conference on Computer Vision and
  Pattern Recognition Workshops}, pages 25--32, 2016.

\bibitem{rozsa2016facial}
Andras Rozsa, Manuel G{\"u}nther, Ethan~M Rudd, and Terrance~E Boult.
\newblock Are facial attributes adversarially robust?
\newblock In {\em Pattern Recognition (ICPR), 2016 23rd International
  Conference on}, pages 3121--3127. IEEE, 2016.

\bibitem{rozsa2017facial}
Andras Rozsa, Manuel G{\"u}nther, Ethan~M Rudd, and Terrance~E Boult.
\newblock Facial attributes: Accuracy and adversarial robustness.
\newblock {\em Pattern Recognition Letters}, 2017.

\bibitem{papernot2016limitations}
Nicolas Papernot, Patrick McDaniel, Somesh Jha, Matt Fredrikson, Z~Berkay
  Celik, and Ananthram Swami.
\newblock The limitations of deep learning in adversarial settings.
\newblock In {\em Security and Privacy (EuroS\&P), 2016 IEEE European Symposium
  on}, pages 372--387. IEEE, 2016.

\bibitem{carlini2017adversarial}
Nicholas Carlini and David Wagner.
\newblock Adversarial examples are not easily detected: Bypassing ten detection
  methods.
\newblock In {\em Proceedings of the 10th ACM Workshop on Artificial
  Intelligence and Security}, pages 3--14. ACM, 2017.

\bibitem{rasmus2015semi}
Antti Rasmus, Mathias Berglund, Mikko Honkala, Harri Valpola, and Tapani Raiko.
\newblock Semi-supervised learning with ladder networks.
\newblock In {\em Advances in Neural Information Processing Systems}, pages
  3546--3554, 2015.

\bibitem{french1999catastrophic}
Robert~M French.
\newblock Catastrophic forgetting in connectionist networks.
\newblock {\em Trends in cognitive sciences}, 3(4):128--135, 1999.

\end{thebibliography}

\end{document}